\documentclass[useAMS,usenatbib,letterpaper]{mnras}
\usepackage{epsfig,amsmath,amssymb,natbib,mathtools}
\usepackage{aas_macros}

\DeclareRobustCommand{\okina}{%
  \raisebox{\dimexpr\fontcharht\font`A-\height}{%
    \scalebox{0.8}{`}%
  }%
}
\hypersetup{draft}
\begin{document}

\title[Indicator Spectra]{\LARGE Indicator Power Spectra: Surgical Excision of Non-linearities and Covariance Matrices for Counts in Cells}

\author[A. Repp \& I. Szapudi]{Andrew Repp$^{1,2}$\ \& Istv\'an Szapudi$^2$\\$^1$Christian Liberty Academy, 16-675 Milo St., Kea{\okina}au, HI 96749, USA\\$^2$Institute for Astronomy, University of Hawaii, 2680 Woodlawn Drive, Honolulu, HI 96822, USA}

\date{\today; submitted to MNRAS}

\label{firstpage}
\pagerange{\pageref{firstpage}--\pageref{lastpage}}
\maketitle

\begin{abstract}
We here introduce \emph{indicator functions}, which identify regions of a given density in order to characterize the density dependence of clustering.  After a  general introduction to this tool, we show that indicator-function power spectra are biased versions of the linear spectrum on large scales. We provide a calculation from first principles for this bias, we show that it reproduces simulation results, and we provide a simple functional form for the translinear portion of the indicator-function spectra. We also outline two applications: first, these spectra facilitate surgical excision of non-linearity and thus significantly increase the reach of linear theory. Second, indicator-function spectra permit calculation of theoretical covariance matrices for counts-in-cells (CIC), facilitating parameter estimation with complementary CIC methods. 
\end{abstract}


\section{Introduction}
\label{sec:intro}

$\Lambda$CDM provides a concordance framework for understanding the evolution of the Universe. It yields a good fit to all measurements to date of the cosmic microwave background (CMB) (e.g., \citealp{Planck2018}) and of galaxy statistics (e.g., \citealp{Salam2017,Elvin-Poole2018}). Further refinements -- whether tighter constraints on $\Lambda$CDM parameters or evidence for extensions and alternatives -- will come from future surveys, both of the CMB (e.g., CMB-S4, \citealp{CMB-S4}) and of the galaxy distribution  (e.g., \emph{Euclid}, \citealp{Euclid}; and \emph{Roman}, \citealp{WFIRST}).

The standard tool for analysis of such data is the two-point correlation function or its Fourier analogue, the power spectrum. For the matter power spectrum $P(k)$, linear theory \citep{Peebles1980, KodamaSasaki} provides a prescription $P_\mathrm{lin}(k)$, with per cent level accuracy\footnote{Based on the implementations in CAMB (Code for Anisotropies in the Microwave Background, \citealp{CAMB}), http://camb.info/} up to $k = .092 h$ Mpc$^{-1}$ and five per cent accuracy up to $k = 0.12 h$ Mpc$^{-1}$.

However, \emph{Euclid} and \emph{Roman} will probe much higher wavenumbers (e.g., \citealp{Agarwal2021} estimate $k_\mathrm{max} = 0.35h$ and $0.60h$ Mpc$^{-1}$, respectively);\footnote{\citet{Agarwal2021} define $k_\mathrm{max}$ as the value at which theoretical systematics exceed $0.25\sigma$ for a parameter. The smaller volume surveyed by \emph{Roman} translates into greater parameter uncertainty and thus a higher value for $k_\mathrm{max}$.} these wavenumbers are well outside the linear regime. Although we have in hand accurate prescriptions for the non-linear spectrum \citep{Smith_et_al, Takahashi2012}, in this regime the information content no longer scales as $k^3$, a fact due to beat coupling \citep{Hamilton2006}, the correlation of Fourier modes \citep{MeiksinWhite1999}, and the highly-tailed, non-Gaussian character of the matter distribution \citep{Carron2011, Repp2015}. As a consequence, there is little independent information in the translinear modes of the power spectrum \citep{RimesHamilton2005,Neyrinck2006}, and much of the information escapes the entire moment hierarchy, becoming inaccessible to perturbative methods \citep{CarronNeyrinck2012}. The result is an ``information plateau,'' reflecting the fact that on these non-linear scales the power spectrum no longer efficiently captures the information in the data \citep{NeyrinckSzapudi2007,LeePen2008,WCS2015,Wolk2015}.

One method of circumventing this plateau relies on the logarithmic transform \citep{NSS09}, and in particular on the fact that the log power spectrum recovers this lost information \citep{CarronSzapudi2013}. \citet{Repp2017} provide a prescription for the log spectrum (for near-concordance cosmologies), the applicability of which one can generalize to discrete tracers like galaxies \citep{CarronSzapudi2014,ReppSzapudi2018,ReppSzapudi2019}. Application of these prescriptions can potentially rescue the bulk of the information lost from the power spectrum.

A second approach is to use counts-in-cells (CIC) jointly with the power spectrum (e.g., \citealp{Gruen2018,ReppSDSSLin}). One barrier to this approach has been the lack of a prescription for the CIC covariance matrix, but \citet{ReppSzapudi2021} show that one can derive this matrix from the volume-averaged correlations of indicator functions.\footnote{Denoted ``slice fields'' in \citet{ReppSzapudi2021}; see below for definitions.} It is possible to measure these averages from the data, but noise in the measurement adversely affects the usefulness of the resulting covariance matrices.

A third approach, is to consider the density dependence of clustering. Examples of this approach include the use of marked correlation functions, in which the weight of a pair depends on properties of each point (e.g., \citealp{Beisbart2000,Szapudi2000b,Skibba2006}), since the use of densities as marks (e.g., \citealp{White2009,White2016,Massara2021}) allows separate analysis of high- and low-density regions. Such functions are inherently four-point statistics (two points and two spatially-varying marks). Other examples include separate analysis of voids and clusters  (e.g. \citealp{Granett2008,Mao2017,Contarini2021}); research into and measurement of environment-dependent correlations (e.g., \citealp{Abbas2005,Alam2019}); clipping (e.g., \citealp{Simpson2016}); and position-dependent power spectra (e.g., \citealp{Chiang2014}). In addition, the sliced correlation functions of \citet{NeyrinckSzapudi2018} are closely related to the technique we develop below.

With this in mind, we here introduce indicator functions as a tool for cosmological analysis. Section~\ref{sec:indfcns} defines these functions, which isolate locations of given density. We begin characterization of their power spectra in Section~\ref{sec:bias}, showing that on large scales they are biased forms of the linear spectrum $P_\mathrm{lin}(k)$; we derive an expression for this bias and show that it reproduces simulation results. In Section~\ref{sec:fits} we provide a simple fit for the remainder of the indicator-function spectrum.

We then outline two applications of indicator-function power spectra. First (consonant with the third approach above), Section~\ref{sec:infogain} notes that departure from linearity depends on density; thus for each density bin we can employ linear theory as far as possible in that particular bin, without undue concern for the linearity of the whole. In this way one can surgically remove as much non-linearity as possible, achieving maximal applicability of linear theory.

Second (consonant with the second approach above), Section~\ref{sec:covmatr} shows that this bias prescription allows a vast reduction of noise in the calculation of counts-in-cells covariance matrices. Thus it facilitates a significant increase in the use of CIC techniques. Section~\ref{sec:concl} summarizes and outlines further applications for future work.

\section{Indicator Functions}
\label{sec:indfcns}

\begin{figure}
    \leavevmode
    \epsfxsize=8.5cm
    \epsfbox{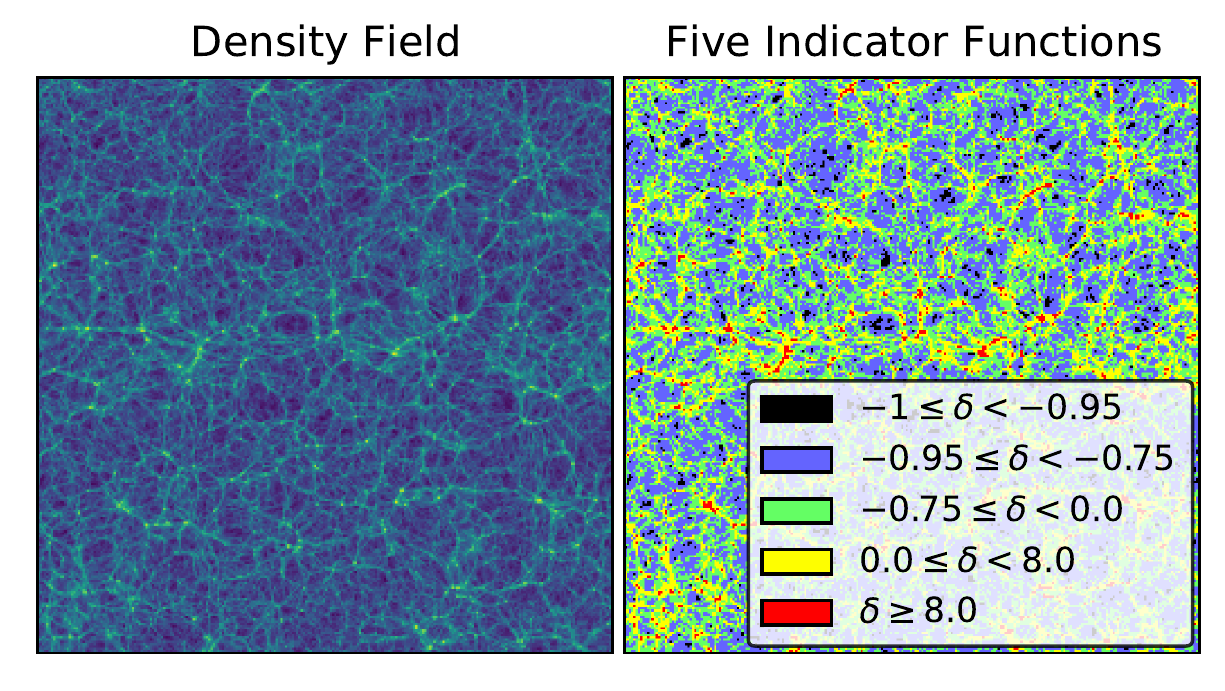}
    \caption{Left panel: the density field of a slice (one pixel thick) through the Millennium Simulation, with density computed in cubical pixels of side length $1.95\mathit{h}^{-1}$Mpc. Right panel: indicator functions (on the same slice) for five density bins, with each color marking the locations at which the corresponding indicator function is non-zero.}
\label{fig:indfcns}
\end{figure}

We borrow the term \emph{indicator function} from mathematics to denote a mapping from a set $X$ to $\{0,1\}$; in particular, for any $B \subseteq X$, its indicator function $\mathcal{I}_B$ takes a value of 1 on the subset $B$ and 0 elsewhere. Thus
\begin{equation}
\mathcal{I}_B(x) = \begin{cases} 
      1 & x \in B \\
      0 & \mbox{otherwise}.
   \end{cases}
\end{equation}

Now, for $X$ we take the overdensity field $\delta(\mathbf{r}) = \rho(\mathbf{r})/\overline{\rho} - 1$, where $\rho$ is the density at a given point $\mathbf{r}$ (or survey cell $\mathbf{r}_i$). Then for any bin $B$ of densities $\delta$, the indicator function $\mathcal{I}_B$ has a value of 1 at all points for which $\delta$ is in the bin $B$; at all other points, $\mathcal{I}_B$ vanishes. (See Figure~\ref{fig:indfcns} for a visual representation of these functions on simulation data.)

Note that indicator functions as here defined are identical to the ``slice fields'' of \citet{ReppSzapudi2021}; the present terminology both recognizes its mathematical usage and avoids confusion with the sliced correlation functions of \citet{NeyrinckSzapudi2018}. One can recover these sliced correlation functions by performing a 1-point average of the indicator-function correlations.

\section{Indicator Function Bias}
\label{sec:bias}

\subsection{First-Order Bias Prescription}
\label{sec:biasprescrip}

As \citeauthor{ReppSzapudi2021} note (\citeyear{ReppSzapudi2021}), the analogy to \citet{Kaiser1984} galaxy bias suggests that one can reasonably expect the large-scale indicator-function spectra to be biased versions of the linear spectrum. Here we show this expectation to be well-founded and derive an expression for this indicator-function bias.

We begin with the assumption of a roughly lognormal matter distribution, so that the distribution of $A = \ln(1+\delta)$ is approximately Gaussian.\footnote{\citet{ReppApdf} provide a prescription for the more precise -- but less tractable -- Generalized Extreme Value distribution for $A$.} (\citealt{BernardeauKofman1995} show that the lognormal distribution -- for a typical power spectrum slope and moderate $\sigma$ -- yields results similar to those of perturbation theory.)

Now let $B_1$ and $B_2$ be two density bins, and let $\mathcal{I}_1$ and $\mathcal{I}_2$ be the corresponding indicator functions. Then the probability $\mathcal{P}(B_1)$ that the density at a given location falls within $B_1$ is simply $\langle \mathcal{I}_1\rangle$; likewise $\mathcal{P}(B_2) = \langle \mathcal{I}_2 \rangle$.

Now for two locations separated by distance $r$, we have
\begin{equation}
\langle \mathcal{I}_1\mathcal{I}_2 \rangle_r = \mathcal{P}(B_1) \mathcal{P}(B_2) (1 + \xi_{12}(r)).\label{eq:b4norm}
\end{equation}
Here $\xi_{12}$ is the cross-correlation function of the normalized indicator functions $\delta_{\mathcal{I}_1}$ and $\delta_{\mathcal{I}_2}$, with $\delta_{\mathcal{I}_i} \equiv \mathcal{I}_i/\langle\mathcal{I}_i\rangle - 1 = \mathcal{I}_i/\mathcal{P}(B_i) - 1$. (In the remainder of this work, any reference to indicator-function correlations, power spectra, etc. always assumes this normalization of the indicator functions.)

Given a roughly Gaussian distribution for $A$, we further normalize to $\nu \equiv (A - \overline{A})/\sigma_A$. Then the left-hand side of Equation~\ref{eq:b4norm} is
\begin{equation}
\langle \mathcal{I}_1 \mathcal{I}_2 \rangle_r = \int d\nu_1\,d\nu_2 \,W_{\!B_1}(\nu_1) W_{\!B_2}(\nu_2) \mathcal{P}(\nu_1,\nu_2),\label{eq:expvalint}
\end{equation}
where $W_{B}(\nu) = 1$ if $\nu$ is within the bin $B$ (and vanishes otherwise).

To write the joint probability explicitly, we use the covariance matrix for $\nu_1$, $\nu_2$ which is
\begin{equation}
C = 
\begin{bmatrix}
1 & \gamma \\
\gamma & 1
\end{bmatrix},
\end{equation}
where $\gamma = \xi_A(r)/\sigma_A^2$. Then
\begin{align}
\mathcal{P}(\nu_1,\nu_2) & = \frac{1}{2\pi\sqrt{|C|}} \exp \left( -\frac{1}{2} \mathbf{v}^\mathrm{T} C^{-1} \mathbf{v} \right)\\
  & = \frac{1}{2\pi\sqrt{1-\gamma^2}} \exp \left(-\frac{\nu_1^2 + \nu_2^2 - 2\gamma\nu_1\nu_2}{2(1-\gamma^2)} \right).\label{eq:jntprb}
\end{align}

For the Millennium Simulation smoothed on $2h^{-1}$-Mpc scales, $\gamma < 0.01$ for $r > 30h^{-1}$Mpc, corresponding to $k \la 0.2h$ Mpc$^{-1}$. So, in this regime we assume that $\xi_A(r) \ll \sigma_A^2$ and expand to first order in $\gamma$:
\begin{align}
\mathcal{P}(\nu_1,\nu_2) & = \frac{1 + \gamma\nu_1\nu_2}{2\pi} \exp \left(-\frac{\nu_1^2 + \nu_2^2}{2}\right)\\
                                        & = (1 + \gamma\nu_1\nu_2) \mathcal{P}(\nu_1)\mathcal{P}(\nu_2)\label{eq:O1jntprb}.
\end{align}
Writing $\mathcal{P}_i$ for $\mathcal{P}(B_i)$, it follows that
\begin{align}
\MoveEqLeft \mathcal{P}_1\mathcal{P}_2 ( 1 + \xi_{12}(r)) \nonumber\\
 &= \int d\nu_1\,d\nu_2 \,W_{\!B_1}(\nu_1) W_{\!B_2}(\nu_2) (1 + \gamma\nu_1\nu_2) \mathcal{P}(\nu_1)\mathcal{P}(\nu_2)\\
 &= \mathcal{P}_1\mathcal{P}_2 \left(1 + \gamma \langle \nu \rangle_{\!B_1} \langle \nu \rangle_{\!B_2} \right).
\end{align}
We conclude that
\begin{equation}
\xi_{12}(r) = \xi_A(r) \frac{(\langle A \rangle_{\!B_1} - \overline{A})(\langle A \rangle_{\!B_2} - \overline{A})}{\sigma_A^4} = \xi_A(r) \frac{\langle\nu\rangle_{\!B_1} \langle\nu\rangle_{\!B_2}}{\sigma_A^2}.\label{eq:O1xiI}
\end{equation}
In particular, for the indicator-function power spectrum for a bin $B$, we set $B_1 = B_2 = B$ so that $P_\mathcal{I}(k) = b_\mathcal{I}^2 P_A(k)$, with
\begin{equation}
b_\mathcal{I}^2 = \frac{(\langle A \rangle_{\!B} - \overline{A})^2}{\sigma_A^4} = \frac{\langle\nu\rangle^2_{\!B}}{\sigma_A^2}.\label{eq:biaswrtA}
\end{equation}
For clarity, note that $\overline{A}$ denotes the average value of $A$ over the entire field, whereas $\langle A \rangle_{\!B}$ is the average value of $A$ over locations with density in the bin $B$.

Equation~\ref{eq:biaswrtA} allows us to calculate the indicator function spectrum $P_\mathcal{I}(k)$ in the linear regime, given the log spectrum $P_A(k)$. \citet{Repp2017} in turn show that $P_A(k)$ is a biased version of the linear power spectrum $P_\mathrm{lin}(k)$, so that
\begin{equation}
P_A(k) = b_A^2 P_\mathrm{lin}(k),
\label{eq:Abias}
\end{equation}
with $b_A^2 = \sigma_A^2/\sigma_\mathrm{lin}^2$.
Thus, the first-order expression for the indicator-function spectrum in the linear regime is
\begin{equation}
P_\mathcal{I}(k) = b_\mathcal{I}^2 b_A^2 P_\mathrm{lin}(k) = \langle\nu\rangle_{\!B}^2 \frac{P_\mathrm{lin}(k)}{\sigma_\mathrm{lin}^2}.
\label{eq:1ordpred}
\end{equation}
Calculation of $\nu$ requires $\overline{A}$ and $\sigma_A^2$, both of which are given by the prescription of \citet{ReppApdf}.\footnote{If the bin is too wide to approximate as a delta function, one can obtain $\langle \nu \rangle_{\!B}$ from the PDF prescription of \citet{ReppApdf}; alternatively, one can measure it from the survey data.}

Note that this bias takes the same form as the \citet{Kaiser1984} galaxy bias, also derived in the small-correlation limit. However, the Kaiser bias assumes $\nu \gg 1$ and applies to a density bin $[\nu, \infty)$, yielding
\begin{equation}
\xi^{\mathrm{gal}}_{>\nu}(r) = \frac{\nu^2}{\sigma^2}\xi(r).
\end{equation}
The indicator bias, on the other hand, applies to an arbitrary (usually narrow) density bin and makes no such assumptions about $\nu$.

We show the two expressions to be consistent, as follows: fix a value $\nu_0$ and consider the density bin $B = \{\nu \,| \,\nu \ge \nu_0\}$. Then, assuming a Gaussian matter distribution as in \citet{Kaiser1984}, Equation~\ref{eq:1ordpred} predicts a bias of $\langle\nu\rangle_B^2 / \sigma^2$. Now
\begin{equation}
\langle\nu\rangle_{\!B}  = \frac{\displaystyle\int_{\nu_0}^\infty d\nu\, \nu e^{-\nu^2\!/2}}{\displaystyle\int_{\nu_0}^\infty d\nu\, e^{-\nu^2\!/2}} = \frac{\displaystyle e^{-\nu_0^2/2}}{\displaystyle \sqrt{\frac{\pi}{2}} \, \mathrm{erfc} \, \frac{\nu_0}{2}};
\end{equation}
using the asymptotic expansion
\begin{equation}
\mathrm{erfc}\,x = \frac{e^{-x^2}}{x\sqrt\pi} + O\left( x^{-3} e^{-x^2} \right),
\end{equation}
it follows that for large $\nu_0$ we have $\langle\nu\rangle_{\!B}  = \nu_0$, reproducing the Kaiser result.

\subsection{Testing the Prescription}
\label{sec:testprescript}

We proceed to test the prescription of Equation~\ref{eq:1ordpred} against the Millennium Simulation \citep{Springel2005},\footnote{https://wwwmpa.mpa-garching.mpg.de/millennium} which consists of a periodic cube with $500h^{-1}$-Mpc sides and approximately $80.6h^3$Mpc$^{-3}$ dark matter particles. We compute density in $256^3$ cubical pixels ($1.95h^{-1}$-Mpc sides). Working at $z = 0$ in the original simulation cosmology, we identify 12 density bins (roughly logarithmic) from $\delta = -0.95$ to 200 ($A = -3$ to 5.3).

\begin{figure*}
    \leavevmode
    \epsfxsize=17cm
    \epsfbox{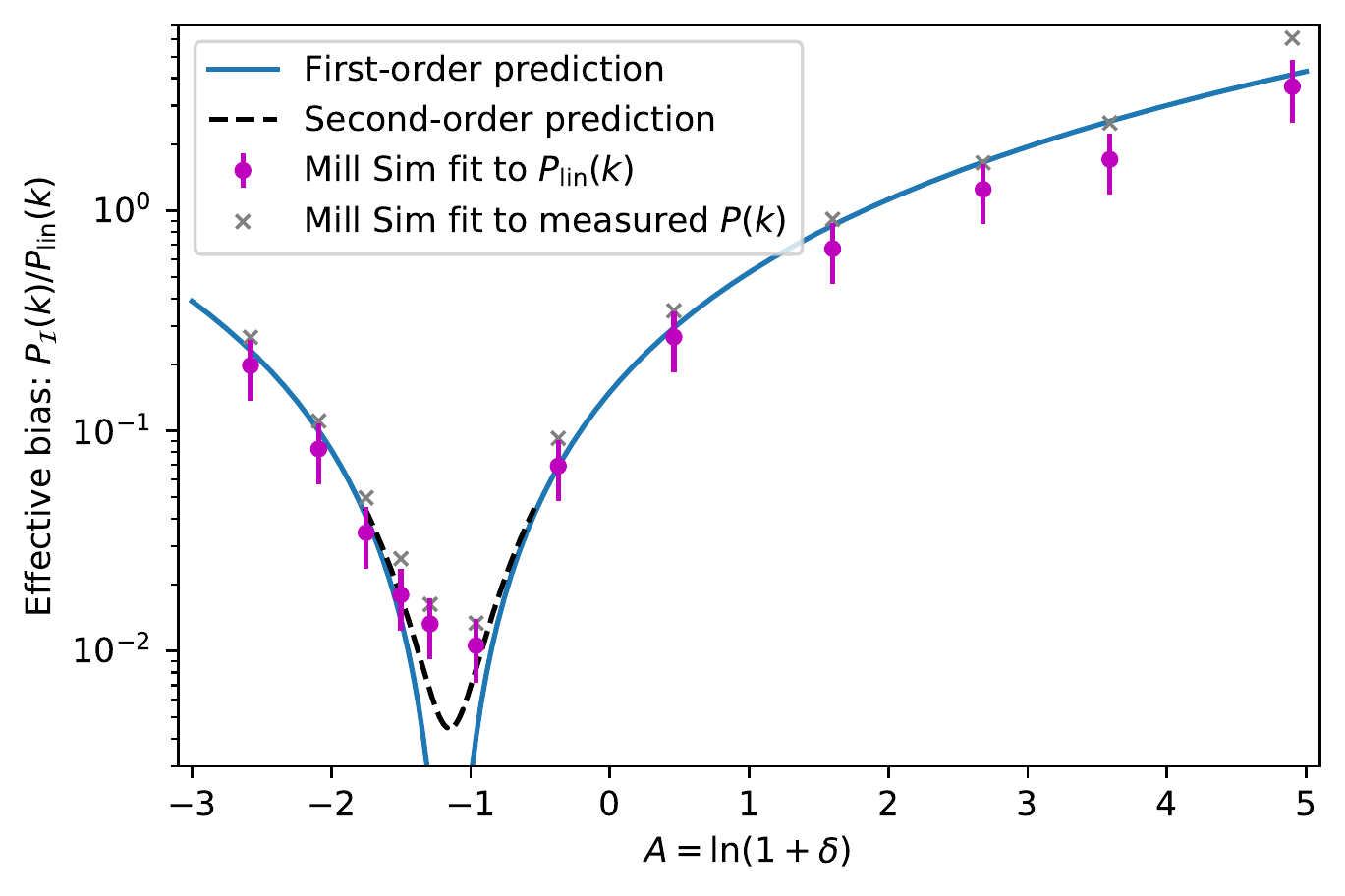}
    \caption{Predicted effective biases (from Equations~\ref{eq:1ordpred} and \ref{eq:2ordpred}), compared with biases measured in the Millennium Simulation. Magenta points -- whose error bars include the effect of cosmic variance (see text) -- show the best fit of Millennium Simulation measurements to the linear power spectrum given by CAMB; grey points show the best fit of Millennium Simulation measurements to the small-$k$ power spectrum measured in that simulation.}
\label{fig:biases}
\end{figure*}

Two issues arise in measuring the indicator spectrum bias from this simulation. First, to restrict the measurement to the linear regime -- and to avoid the shot noise effects of a small number of cells in a bin -- we determine the bias by finding the best-fit value of $b^2_\mathrm{eff}$ in
\begin{equation}
P_\mathcal{I}(k) = b^2_\mathrm{eff} P_\mathrm{lin}(k)\label{eq:linfit}
\end{equation}
for $k \le .07h$ Mpc$^{-1}$. This $b^2_\mathrm{eff}$ is the effective bias plotted in Figure~\ref{fig:biases}.

Second, the Millennium Simulation is small enough to render cosmic variance significant at these wavenumbers. We estimate the effect of this variance on the Millennium Simulation by fitting, as in Equation~\ref{eq:linfit}, the measured power spectrum $P(k)$ to $P_\mathrm{lin}(k)$, again for $k \le .07h$ Mpc$^{-1}$. If the effect of cosmic variance were negligible, one would expect $P(k) = P_\mathrm{lin}(k)$ and thus an effective bias of unity. Instead, we find a best-fit $P(k) =  0.7P_\mathrm{lin}(k)$ in this regime, indicating a 30 per cent effect. Note that it is unclear that the effect would impact all densities equally. Yet, it is reasonable to approximate the effect of cosmic variance as an additional error term, with $\sigma$ equal to 30 per cent  the size of the best-fit value of $b^2_\mathrm{eff}$. This approximation probably underestimates cosmic variance somewhat, since with 68 per cent likelihood the effect is actually less than $1\sigma$. We thus determine the error bars on the magenta points in Figure~\ref{fig:biases} by adding in quadrature the error on the fit of Equation~\ref{eq:linfit} (estimated by the fit routine), and 30 per cent of the best-fit $b^2_\mathrm{eff}$. Overall, underestimating the cosmic variance is conservative when testing a first-principles prescription.

Referring to Figure~\ref{fig:biases}, we observe that the first-order bias prediction of Equation~\ref{eq:1ordpred} matches the measured bias quite well except for densities near $A = -1.15$, the mean value of $A$. In particular, the first order prediction requires the bias to vanish at $A = \overline{A}$, but the measured bias does not display this behavior. The reason becomes apparent when we consider the second-order expansion in the next section.

To check our estimate of cosmic variance, we can calculate the effective bias of the (measured) indicator spectra with respect to the (measured) non-linear power spectrum $P(k)$ at $k \le .07h$ Mpc$^{-1}$. Upon doing so we obtain the grey points in Figure~\ref{fig:biases}. It is not surprising that they are typically within $1\sigma$ of the magenta points, since the magenta error bars already reflect the deviation of the Millennium Simulation from the CAMB prescription. However, the fact that the grey points are systematically offset from the magenta indicates that the relatively small simulation volume is indeed affecting the measurement of $b^2_\mathrm{eff}$ in a cosmic variance-like way.\footnote{The final point in Figure~\ref{fig:biases} ($A\sim 5$) represents pixels near virialization ($100 < \delta < 200$) on the 2-Mpc scale and constituting less than 0.04 per cent of the simulation volume.}

\subsection{Second-Order Bias Prescription}
\label{sec:2ordpresc}

We now consider the failure of the prediction near the mean value of $A$ by returning to Equation~\ref{eq:jntprb} and expanding to second order in $\gamma$, obtaining
\begin{equation}
\mathcal{P}(\nu_1,\nu_2) = \mathcal{P}(\nu_1)\mathcal{P}(\nu_2)\left(1 + \gamma\nu_1\nu_2 + \frac{\gamma^2}{2}(\nu_1^2-1)(\nu_2^2-1)\right).\label{eq:O2jntprb}
\end{equation}
Thus, recalling that $\gamma = \xi_A / \sigma_A^2$, we have for $\nu_1 = \nu_2 = \nu$ the following analogue of Equation~\ref{eq:O1xiI}:
\begin{equation}
\xi_\mathcal{I} = \frac{\langle\nu\rangle^2}{\sigma_A^2} \xi_A + \frac{(1-\langle\nu^2\rangle)^2}{2\sigma_A^4}\xi_A^2,
\label{eq:O2xiI}
\end{equation}
with the average $\langle \cdot \rangle$ taken over the bin $B$ defining $\mathcal{I}$.
Passing to Fourier space and recalling Equation~\ref{eq:Abias},
\begin{equation}
P_\mathcal{I}(k) = \langle\nu\rangle^2 \frac{P_\mathrm{lin}(k)}{\sigma_\mathrm{lin}^2} + \frac{(1-\langle\nu^2\rangle)^2}{2} \frac{P_\mathrm{lin}^{(*2)}(k)}{\sigma_\mathrm{lin}^4},\label{eq:2ordpred}
\end{equation}
where $P_\mathrm{lin}^{(*2)}(k)$ is the self-convolution of the linear power spectrum.

Thus we see that the second-order term contains a constant, with the result that, as $\nu$ approaches zero ($A \longrightarrow \overline{A}$), this second-order term becomes dominant. This fact accounts for the failure of the first-order prediction when $A \approx \overline{A}$.

Calculating the self-convolution is theoretically straightforward but numerically tricky. Accomplishing the full three-dimensional self-convolution of $P_\mathrm{lin}(k)$ is computationally expensive, so we instead perform a Hankel transformation on the theoretical $P_\mathrm{lin}(k)$ (from CAMB) to get $\xi_\mathrm{lin}(r)$; we then square $\xi_\mathrm{lin}(r)$ and perform the inverse transformation to obtain $P_\mathrm{lin}^{(*2)}(k)$. Our algorithm for the transformations employs the implementation by \citet{Szapudi2005} of the quadrature formula of \citet{Ogata2005}. To handle the endpoints of the theoretical predictions, we apodize by extrapolating a constant power law slope.

In this way we obtain the theoretical $P_\mathrm{lin}^{(*2)}(k)$ and thus the dashed curve in Figure~\ref{fig:biases}. We conclude that, for an accurate bias prescription in the linear regime, it is sufficient to use the first-order prediction (Equation~\ref{eq:1ordpred}) for $|\nu| \ga 0.5$; otherwise, the second-order prediction (Equation~\ref{eq:2ordpred}) is necessary. Note that in the case under consideration, $\overline{A} = -1.15$ and $\sigma_A^2 = 1.52$; thus $\nu < 0.5$ corresponds to $\delta < -0.4$, where discrete tracers such as galaxies are typically absent anyway.

\section{Fitting $P_\mathcal{I}(k)$}
\label{sec:fits}

Fig.~\ref{fig:biases} shows that the simple prescription $P_\mathcal{I}(k) \propto P_\mathrm{lin}(k)$, with the bias of Equation~\ref{eq:1ordpred} (or Equation~\ref{eq:2ordpred} for $\nu^2 \la 0.25$), yields accurate results in the linear regime. However, it is worth investigating two extensions that allow us to fit these spectra to higher wavenumbers.

\begin{figure}
    \leavevmode
    \epsfxsize=8.5cm
    \epsfbox{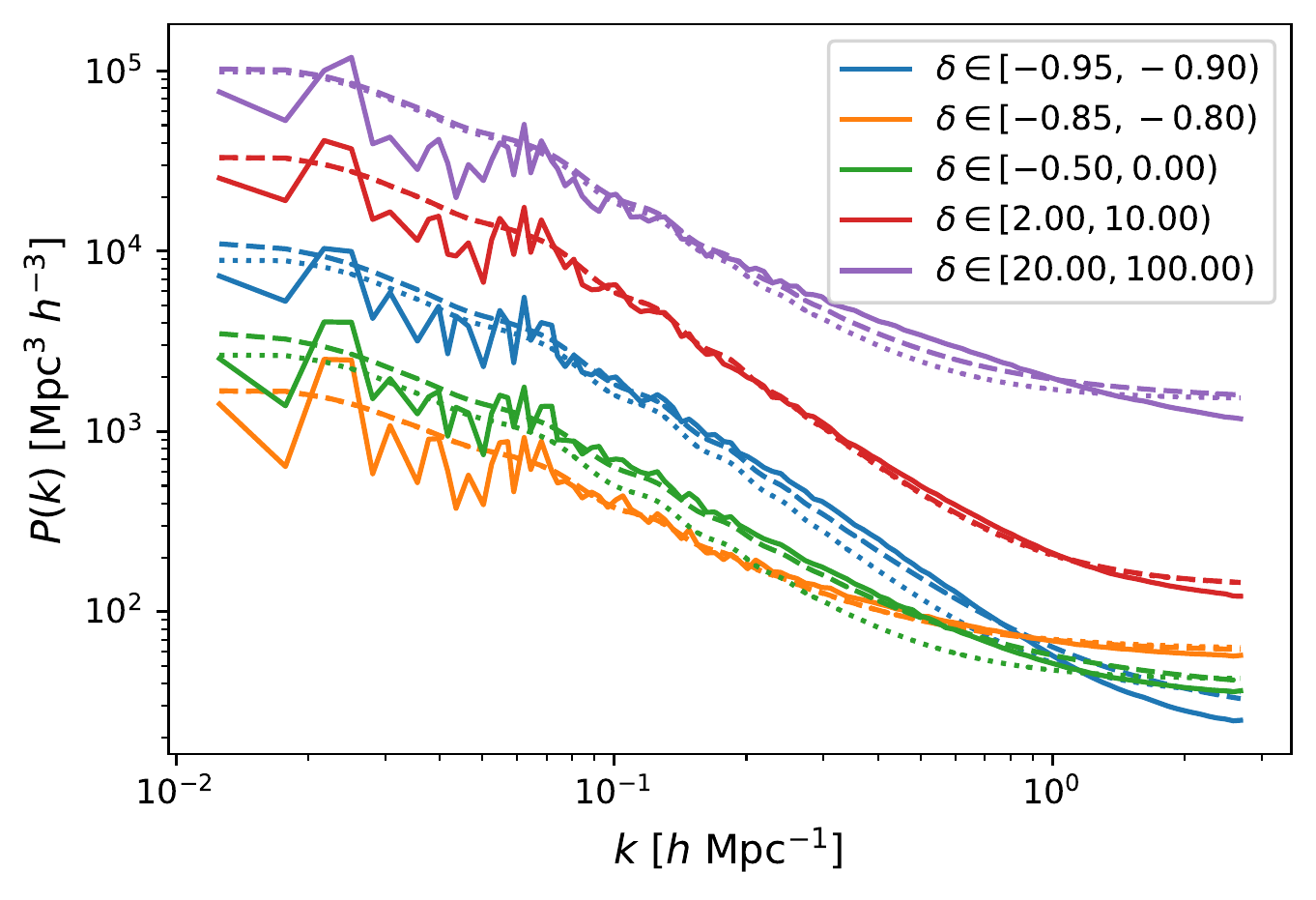}
    \caption{Fits to indicator power spectra: solid lines -- measured spectra; dotted lines -- bias plus constant (Equations~\ref{eq:linmod}--\ref{eq:Clin}); dashed lines -- bias plus power law plus constant (Equation~\ref{eq:pwrlawmod}).}
\label{fig:fitPk}
\end{figure}

The first extension involves the addition of a constant to satisfy an integral constraint. Figure~\ref{fig:fitPk} exhibits a high-$k$ plateau for the measured indicator-function spectra, due to discreteness; however, because the distribution is not Poisson (the only allowed values being 0 and 1) we cannot use the familiar $1/\overline{n}$ formula for shot noise. Nevertheless, we can determine the necessary additive constant from the fact that the integral of the power spectrum equals the variance.

If we consider the indicator function $\mathcal{I}$ for a density bin $B$, then both its first and second moments equal the probability $\mathcal{P}(B)$ that a cell falls within the bin $B$; hence the variance of $\mathcal{I}$ is $\mathcal{P}(B) ( 1 - \mathcal{P}(B))$. It follows that the variance of the normalized indicator function (see immediately after Equation~\ref{eq:b4norm}) is
\begin{equation}
\sigma^2_{\delta_\mathcal{I}} = \frac{1}{\mathcal{P}(B)} - 1.
\end{equation}
If we now model the indicator spectrum as Equation~\ref{eq:1ordpred} plus a constant $C$, we can write
\begin{align}
\frac{1}{\mathcal{P}(B)} - 1 & = \int \frac{d^3k}{(2\pi)^3} \left( \langle\nu\rangle_{\!B}^2 \frac{P_\mathrm{lin}(k)}{\sigma_\mathrm{lin}^2} + C\right)\\
                                          & = \langle\nu\rangle_{\!B}^2 + \frac{C}{\delta V},
\end{align}
where $\delta V$ is the volume of a survey cell. Since we can obtain $\sigma^2_\mathrm{lin}$ from linear theory, we can solve for $C$ to obtain
\begin{gather}
P_\mathcal{I}(k) = \langle\nu\rangle_{\!B}^2 \frac{P_\mathrm{lin}(k)}{\sigma^2_\mathrm{lin}} + C\label{eq:linmod}\\
C = \delta V \left(\frac{1}{\mathcal{P}(B)} - 1 - \langle\nu\rangle_{\!B}^2 \right).
\label{eq:Clin}
\end{gather}

The results of adding this constant appear as dotted curves in Figure~\ref{fig:fitPk}. It is evident that this procedure yields the correct order of magnitude, but the slope of the indicator spectra is often more gradual than this prescription would suggest. However, the advantage of this prescription is that it involves only linear theory plus discreteness, a fact we shall use in Section~\ref{sec:infogain} below.

We now consider a second extension, namely the addition of a power law $Dk^n$, so that our model is as follows:
\begin{equation}
P_\mathcal{I}(k) = \langle\nu\rangle_{\!B}^2 \frac{P_\mathrm{lin}(k)}{\sigma_\mathrm{lin}^2}  + Dk^n + C.\label{eq:pwrlawmod}
\end{equation}
(For bins near $\nu = 0$ we also include a second-order term, as in Equation~\ref{eq:2ordpred}.) Here we calculate the bias term(s) as before; we then fit $D$ and $n$, and we calculate $C$ from the fact that the power spectrum integrates to the variance. The results appear as dashed lines in Figure~\ref{fig:fitPk}, showing that Equation~\ref{eq:pwrlawmod} provides a reasonably accurate model for the indicator spectra up to $k > 2h$ Mpc$^{-1}$, which is well beyond the reach of planned galaxy surveys. Future work will investigate physical reasons for this power law term.

\section{Excising Non-linearity}
\label{sec:infogain}

Let us now revert to the purely linear indicator spectrum model of Equation~\ref{eq:linmod} (the dotted curves in Fig.~\ref{fig:fitPk}). It is evident from the figure that the range of wavenumbers for which this model provides a good fit depends on the density under consideration.

Thus, the applicability of linear theory depends on both wavenumber and density. Traditional methods, which analyse the power spectrum as a whole, must treat all densities at once and thus restrict the use of linear theory to $k \la 0.1h$ Mpc$^{-1}$. On the other hand, alternate strategies such as clipping remove the same high-density regions for all $k$. The use of indicator function spectra gives us the ability to apply linear theory (Equation~\ref{eq:linmod}) to \emph{each} density bin up to the limit of its accuracy in that bin. In other words, it lets us remove non-linearity -- not by amputation in  $k$- or $\delta$-space, but by surgical excision from the $k$-$\delta$ plane -- leaving only the regions in which Equation~\ref{eq:linmod} is accurate.

As a step toward visualizing the information gained by this method, we estimate the applicability of linear theory in each bin as follows. As noted in Section~\ref{sec:intro}, linear theory is applicable at the 5 per cent level up to $k = 0.12h^{-1}$Mpc. Thus our standard for a ``good fit'' is the match between $P_\mathrm{lin}(k)$ and the measured matter power spectrum $P(k)$ up to 0.12, and at most 5 per cent deviation thereafter.\footnote{This somewhat awkward phrasing is due to the existence of fluctuations in the Millennium Simulation power spectrum on linear scales, and the resulting fact that, on these scales, the measured $P(k)$ almost always deviates from $P_\mathrm{lin}(k)$ by more than 5 per cent. Roughly speaking, before $k = 0.12h^{-1}$Mpc the departures are due to cosmic-variance induced fluctuations; afterwards they are due to non-linearity.} Thus we determine, at each wavenumber, the per cent departure of the measured simulation spectrum $P(k)$ from $P_\mathrm{lin}(k)$. For each density bin, we likewise determine the per cent departure of the measured indicator spectrum $P_\mathcal{I}(k)$ from the linear prescription of Equation~\ref{eq:linmod}. To reduce stochasticity arising from the relatively small simulation volume, we smooth both per cent differences with a rolling average 5 $k$-bins wide. Finally, in each density bin we determine the wavenumber $k_\mathrm{cut}$ at which the per cent difference of the measured $P_\mathcal{I}(k)$ from Equation~\ref{eq:linmod} exceeds the per cent difference of the measured $P(k)$ from $P_\mathrm{lin}(k)$ (before $k = 0.12h^{-1}$Mpc) or exceeds 5 per cent (after $k = 0.12h^{-1}$Mpc).

\begin{figure}
    \leavevmode
    \epsfxsize=8.5cm
    \epsfbox{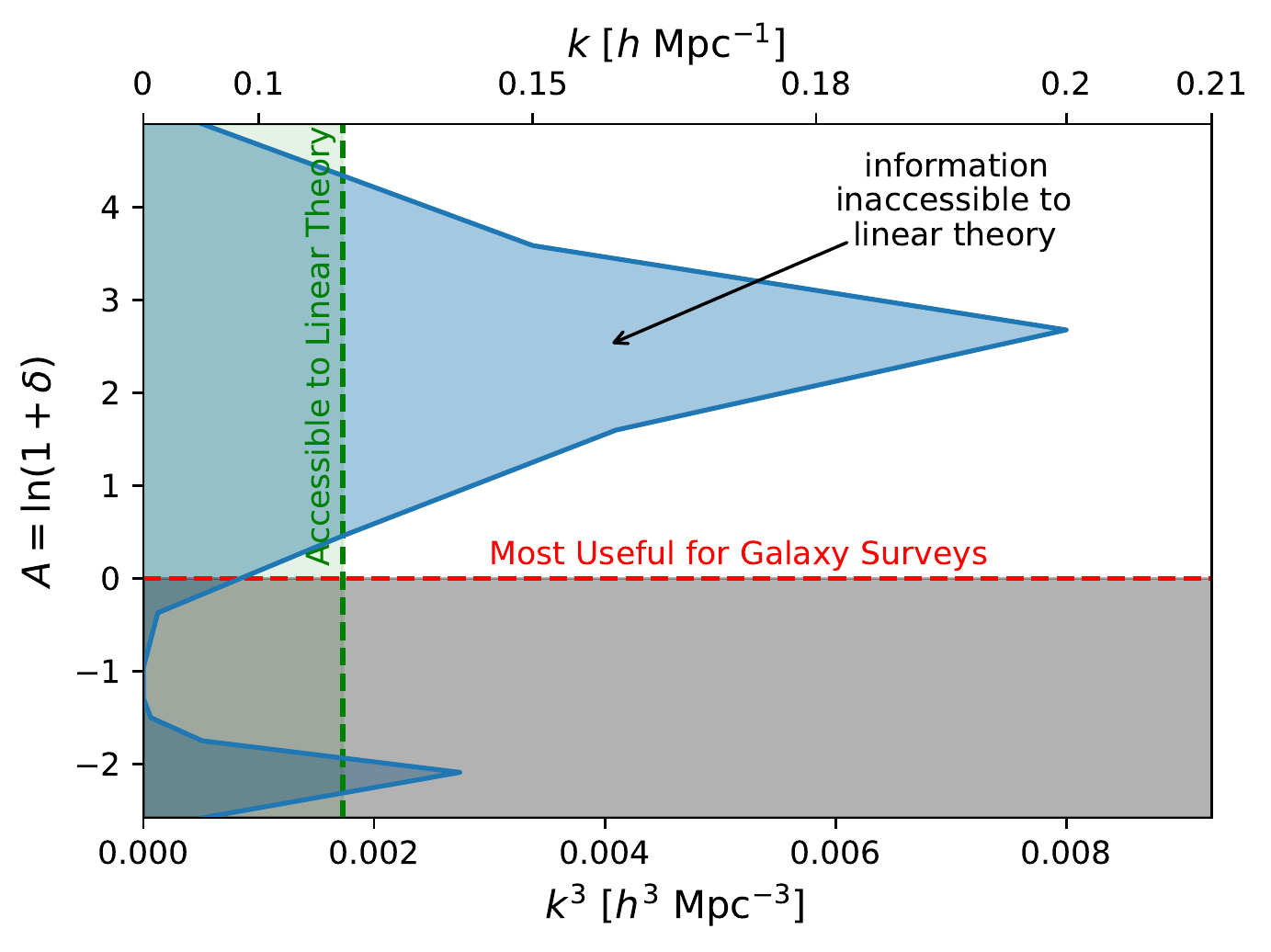}
    \caption{Blue region -- estimated applicability of Equation~\ref{eq:linmod}, which incorporates only linear analysis; green region -- applicability of standard linear theory. Since Fisher information in the linear regime scales roughly as $k^3$, we use this scale on the $x$-axis for a visual impression of the information gain from linear indicator-function analysis.}
\label{fig:linrange}
\end{figure}

We plot these values of $k_\mathrm{cut}$ in Fig.~\ref{fig:linrange}. Since the power-spectrum information content (in the linear regime) scales as the cube of $k$,  we plot the figure as a function of $k^3$ (with $k$ itself displayed on the upper axis), so that area is proportional to information. The figure also displays the 5 per cent applicability limit of linear theory ($k = 0.12h$ Mpc$^{-1}$). Since galaxy surveys have difficulty resolving low-density regions, we mark $\delta = 0$ as an approximate lower limit of the regime ``most useful for galaxy surveys''; in practice this level would depend on the number density achieved by the survey. In any case, it is the upper part of the figure which has the most practical relevance.

The blue region of the figure shows where the indicator spectra are linear. Note that the shape of the blue region roughly conforms to expectations from Equation~\ref{eq:O2xiI}. This equation implies two situations in which the dominance of the second-order term would cause linear predictions to break down. First, for small $\nu$ (i.e., $A \sim \overline{A}$) the constant in the second-order term dominates over $\nu^2$ in the first-order term; the resulting failure of linearity at $A \approx -1$ is evident in Figure~\ref{fig:linrange}. Second, for large $|\nu|$, the second-order $\nu^4$ again dominates over the first-order $\nu^2$, causing the breakdown of linearity notable at the top and bottom of the plot.

It is likely that, in this region, the information continues to scale as $k^3$ and that the covariance matrix (in $B$ for various cosmological parameters) is at least block-diagonal. However, the potential for cross-correlations complicates the analysis. Quantification of this information gain requires computation of the Fisher matrix and is beyond the scope of this work; nevertheless -- and despite the breakdowns of linearity noted above -- the figure visually demonstrates that the gain will be substantial.

A few caveats are appropriate at this point. First, the $k^3$ scaling assumes, for a parameter $\alpha$, that the derivative $\partial \ln P_\mathrm{lin}(k)/\partial \alpha$ is constant in $k$ -- i.e., that the parameter $\alpha$ is amplitude-like. Fortunately, some of the parameters of greatest interest (such as $w$, $w_a$, and $\sigma_8$) primarily affect the power spectrum amplitude on linear scales. Second, the scaling assumes that marginalization over the parameter covariances does not wipe out the information gained on a particular parameter $\alpha$. In such a case, breaking the degeneracy requires resort to other methods; for instance, use of counts-in-cells (CIC) can break the degeneracy between $\sigma_8$ and galaxy bias \citep{ReppSDSSLin}. Indicator functions can facilitate the application of some of these methods -- see Section~\ref{sec:covmatr} for their use in deriving CIC covariance matrices. However, in this section we focus primarily on the extension of linear theory to smaller scales and thus do not address other built-in limitations of linear theory \emph{per se} (such as degeneracies).

Furthermore, various perturbative models provide accurate descriptions of the power spectrum beyond $k = 0.12h$ Mpc$^{-1}$. It would in principle be possible -- and certainly desirable -- to apply perturbation theory to indicator functions and thus to extend their use to mildly non-linear scales. In general, models for indicator spectra at smaller scales would be highly useful; Equation~\ref{eq:pwrlawmod} represents a first -- albeit phenomenological and non-perturbative -- attempt at such modeling.

We note also that the ability to access smaller scales with linear theory is only one aspect of indicator-function utility, since one can reconstruct, via linear combination, the log power spectrum from indicator spectra. Since the log transform (in large part) gaussianizes the matter distribution, analysis of the log spectrum can extract essentially all cosmological information contained in the data (see \citealp{CarronSzapudi2013} for demonstration and caveats). It follows that the set of indicator spectra contains at least as much information as the log spectrum, which in turn (at reasonable non-linear scales) contains significantly more information than the power spectrum itself. We thus submit that it is worthwhile to investigate the application of indicator-function methods to both perturbative and emulator-based approaches.

In any case, we conclude that indicator functions provide a means for precise removal of non-linearity, and thus they promise a significant increase in the reach of  linear theory for cosmological parameter estimation.

\section{Theoretical Covariance Matrices}
\label{sec:covmatr}

Analysis of counts-in-cells (CIC) provides a complementary means of obtaining information beyond the linear regime; for instance, use of CIC breaks the degeneracy between $\sigma_8$ and galaxy bias in SDSS data \citep{ReppSDSSLin}. This method requires knowledge of the CIC covariance matrix, but \citet{ReppSzapudi2021} show that one can derive the matrix from volume-averaged indicator correlation functions, as follows: if $\mathcal{P}$ is the probability of a density falling within bin $B$, then the variance of that probability is
\begin{equation}
\sigma^2_{\mathcal{P}} = \frac{\mathcal{P}\left(1-\mathcal{P}\right)}{N_c} + \frac{(N_c - 1) \overline{\xi}_\mathcal{I}^{\neq}}{N_c} \mathcal{P}^2,\label{eq:corrvar}
\end{equation}
where $N_c$ is the number of survey cells. In Equation~\ref{eq:corrvar}, $\overline{\xi}_\mathcal{I}^{\neq}$ is the volume-averaged indicator correlation function (for density bin $B$), excluding zero-lag (intracellular) correlations. Likewise, for covariances, we let $\mathcal{P}_1$, $\mathcal{P}_2$ be the probabilities of a cell falling within (disjoint) density bins $B_1$, $B_2$, respectively; then the covariance of $\mathcal{P}_1$ and $\mathcal{P}_2$ is
\begin{equation}
\sigma_{\mathcal{P}_1\mathcal{P}_2} = \frac{-\mathcal{P}_1 \mathcal{P}_2}{N_c} + \frac{N_c-1}{N_c} \overline{\xi}_{\mathcal{I}_1\mathcal{I}_2} \mathcal{P}_1 \mathcal{P}_2,
\label{eq:covar}
\end{equation}
where $\overline{\xi}_{\mathcal{I}_1\mathcal{I}_2}$ is the volume-averaged indicator cross-correlation function (for bins $B_1$ and $B_2$).

Thus, we need volume-averaged indicator correlation functions to calculate CIC covariance matrices. \citet{ReppSzapudi2021} perform the average using Monte Carlo sampling with indicator-correlation functions measured directly from the data. Unfortunately, this measurement introduces significant noise into the resulting matrix. We here demonstrate that Equation~\ref{eq:1ordpred} (or, near $\overline{A}$, Equation~\ref{eq:2ordpred}) provides a superior alternative.

We begin with the volume-averaged correlation function
\begin{equation}
\overline{\xi} = \frac{1}{V^2} \int d^3r_1\, d^3r_2 \, \xi(\mathbf{r_1},\mathbf{r_2}) W(\mathbf{r_1}) W(\mathbf{r_2}),
\end{equation}
where $W(\mathbf{r})$ is the window function for the survey (unity within the survey volume and zero without). Passing to Fourier space, we obtain
\begin{equation}
\overline{\xi} = \frac{1}{V^2} \int \frac{d^3k}{(2\pi)^3} P(k) |W(\mathbf{k})|^2.
\end{equation}
for an isotropic spectrum.

For a cubical survey (like the Millennium Simulation), we have $W(\mathbf{r}) = W(x)W(y)W(z)$; if $L$ is the survey side length,
\begin{equation}
\frac{W(\mathbf{k})}{V} = \prod_{i=1}^3 \frac{\sin k_i L/2}{k_i L/2},
\end{equation}
so that
\begin{equation}
\overline{\xi} _\mathcal{I}= \int \frac{d^3k}{(2\pi)^3} P_\mathcal{I}(k) \prod_{i=1}^3 \mathrm{sinc}^2\! \left( \frac{k_iL}{2} \right).
\label{eq:xibar}
\end{equation}
Since the function $\mathrm{sinc}(k_iL/2)$ cuts off quickly for $k \ga 2/L$ ($\sim .01h$ Mpc$^{-1}$ for the Millennium), the linear portion of the spectrum exerts the greatest influence on $\overline{\xi}$. Thus analytic methods for calculating $\overline{\xi}_\mathcal{I}$ require a prescription for the indicator spectrum bias, such as that provided by Equation~\ref{eq:1ordpred}.

\begin{figure*}
    \leavevmode
    \epsfxsize=17cm
    \epsfbox{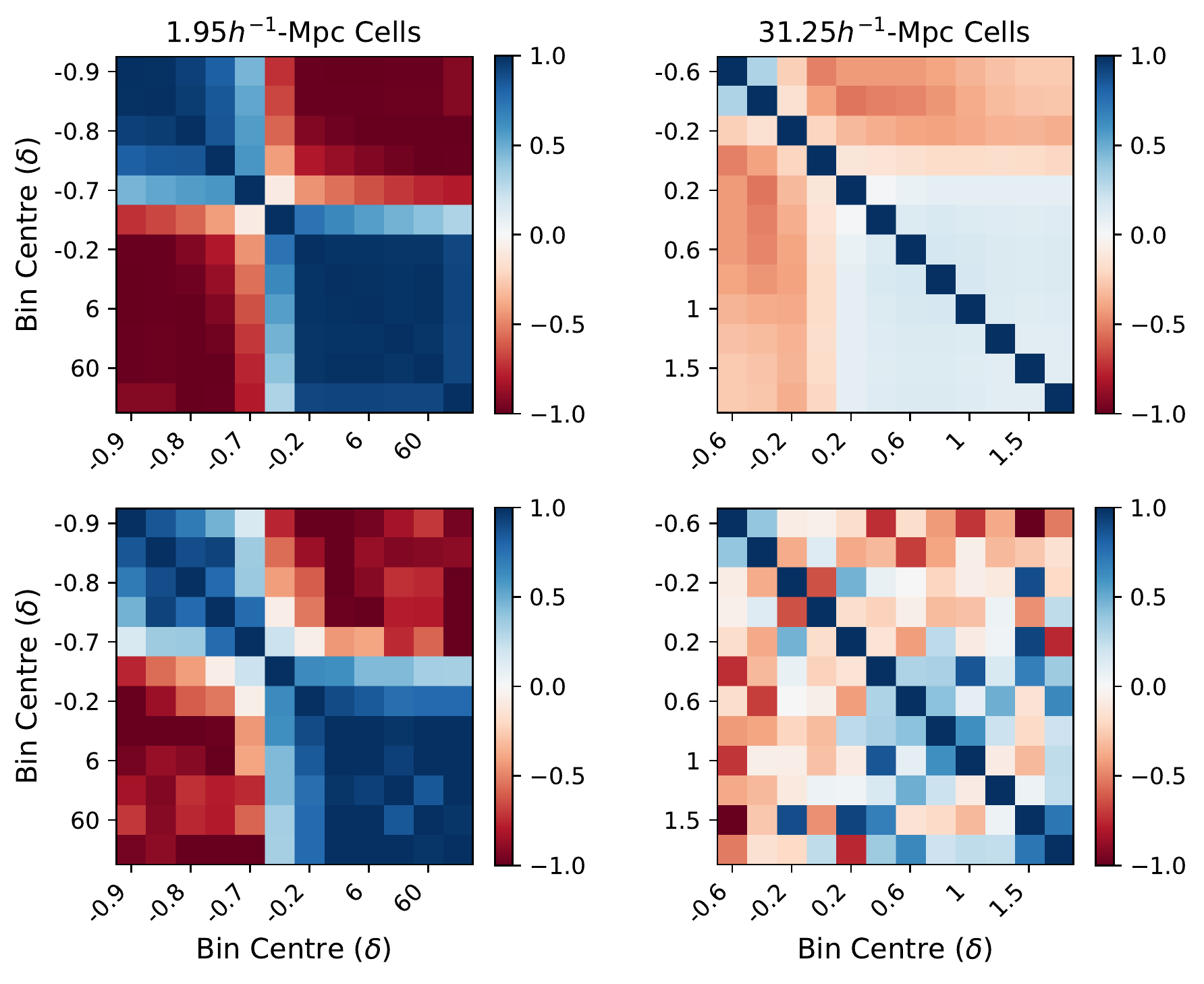}
    \caption{Correlation matrices ($\rho_{ij} = \sigma_{ij}/(\sigma_i \sigma_j)$) for Millennium Simulation counts-in-cells, at two smoothing scales (left panels: $1.95h^{-1}$Mpc; right panels: $31.25h^{-1}$Mpc). Upper panels show matrices calculated from indicator-function spectra, as described in Section~\ref{sec:covmatr} and Appendix~\ref{app:cov}; lower panels show the results of estimating $\overline{\xi}_\mathcal{I}$ from Monte Carlo sampling (as in \citealt{ReppSzapudi2021}) and then calculating the matrices. See text for discussion of the differences between the upper and lower panels.}
\label{fig:covmatr}
\end{figure*}

Diagonal elements require $\overline{\xi}^{\neq}$ rather than $\overline{\xi}$ (Equation~\ref{eq:corrvar}). If $N_c$ is the number of survey cells, we have for an indicator function $\mathcal{I}$
\begin{align}
N_C^2 \, \overline{\xi}_\mathcal{I} & = (N_c^2 - N_c) \overline{\xi}_\mathcal{I}^{\neq}  + N_c \,\xi_\mathcal{I}(0) \\
\overline{\xi}_\mathcal{I}^{\neq} & = \frac{1}{N_c - 1} ( N_c \,\overline{\xi}_\mathcal{I} - \sigma^2_\mathcal{I}) \\
 & = \frac{1}{N_c - 1} \left( N_c \,\overline{\xi}_\mathcal{I} - \frac{1}{\mathcal{P}} + 1 \right).
 \label{eq:xibarneq}
 \end{align}
 
 Hence, the indicator spectrum prescriptions in this work, together with Equations~\ref{eq:xibar} and \ref{eq:xibarneq}, allow us to calculate $\overline{\xi}^{\neq}$ from first principles; thence Equations~\ref{eq:corrvar} and \ref{eq:covar} give us the elements of the CIC covariance matrix.
 
As a proof of concept, we thus calculate the CIC covariance matrices, for two smoothing scales ($1.95h^{-1}$ $31.25h^{-1}$Mpc), for the Millennium Simulation dark matter particles at $z = 0$. (See Appendix~\ref{app:cov} for details of the calculation.) We then normalize to obtain the correlation matrices ($\rho_{ij} = \sigma_{ij}/(\sigma_i \sigma_j)$), which we display on the top row of Fig.~\ref{fig:covmatr}. We see that at $\sim 30 h^{-1}$Mpc the matrix is approximately diagonal; however, even at this scale we note some significant off-diagonal correlation, the strongest being $-0.53$ (between the bins at $\delta = -0.4$ and 0.2).

At $\sim 2h^{-1}$Mpc, however, the correlation matrix is decidedly non-diagonal: 60 per cent of the off-diagonal elements have absolute value exceeding 0.9. The counts at this scale thus display significant correlation; if we interpret the visible blocks as clusters and voids, then the cluster counts are well-correlated, and the void counts are well-correlated, while the two are anti-correlated with each other. (A similar but much weaker effect is evident in the $30h^{-1}$-Mpc case.) At $2h^{-1}$Mpc, the assumption of diagonality in CIC analysis would lead to significantly erroneous results. Thus at this scale, without such knowledge of the covariance matrix one cannot properly compare predicted CIC to those observed in surveys.
 
As noted above, it is possible to measure $\overline{\xi}_\mathcal{I}^{\neq}$ from the data by Monte Carlo sampling, as in \citet{ReppSzapudi2021}; this procedure yields the bottom panels of Fig.~\ref{fig:covmatr}. Comparison of the top and bottom panels demonstrates the superiority of indicator function methods over Monte Carlo sampling (besides a significantly reduced computational load). In particular, the indicator spectrum prescription greatly reduces the noise in the results; this effect is most evident for the right-hand panels ($\sim 30h^{-1}$Mpc), where the noise level reflects the relatively small number of survey cells ($16^3$). In addition, the MC procedure occasionally produces ``correlations'' with absolute value up to 40 per cent greater than unity (in Fig.~\ref{fig:covmatr} we truncate the color range to mathematically meaningful values) -- implying that the covariance matrices from this procedure are internally inconsistent. Even if the effect is strictly numerical, it illustrates the difficulty of obtaining accurate results from this MC method.

Therefore, indicator functions provide a means for calculating the CIC covariance matrix, facilitating application of CIC methods (complementary to linear theory) for parameter estimation.

\section{Summary and Discussion}
\label{sec:concl}

Indicator functions are defined by density bins, such that the function has a value of one at all locations within that bin, vanishing everywhere else. Thus the indicator functions isolate regions of given densities. We here have shown that on large scales the indicator-function spectra are biased versions of the linear power spectrum, and we provide a first-principles prescription for this bias as well as a simple form for fitting the remainder of the spectrum. 

We have also outlined two applications of these functions, both relevant to cosmological parameter estimation. The first follows from the fact that linear indicator-function theory (i.e., biased $P_\mathrm{lin}(k)$ plus a constant) is often valid well past the traditional linear regime. Thus, consideration of indicator-function spectra allows us to excise the densities and wavenumbers affected by non-linearity; it follows that these functions promise to extend to reach of linear theory and significantly increase its capability of extracting cosmological information from survey data. The second is that this linear prescription (bias plus constant) allows calculation of theoretical counts-in-cells covariance matrices and thus facilitates extraction of complementary information using CIC methods.

Furthermore, we note for future work the following potential applications of indicator functions. First, Fig.~\ref{fig:indfcns} seems to show three density regimes characterized by different topologies. Roughly speaking, these three regimes are $\nu \ll 0$ (voids), $\nu \sim 0$ (filaments), and $\nu \gg 0$ (clusters); one could render this delineation more precise by investigating indicator-function percolation thresholds. It is interesting that filaments ($\nu \sim 0$) at the 2-Mpc scale actually correspond to underdensities. This fact reflects the highly-skewed shape of the matter distribution on such scales. One salutary effect of the log transform is the drastic reduction of such skewness, allowing treatment of clusters and voids on a more nearly equal footing.

Second, the power law term in the fit of Equation~\ref{eq:pwrlawmod} warrants further investigation. Identification of its physical basis could lead to a prescription or its replacement by a more accurate form. We note that the analysis of Section~\ref{sec:bias} does not immediately account for the halo/void structure of the small-scale matter distribution. To a first approximation, one would expect indicator functions within a given halo (void) to take the form of spherical shells, and it is tempting to speculate that this 1-halo (void) behavior might affect the high-$k$ shape of the spectra.

Third, characterization of indicator-function spectra would facilitate a first-principles prediction of both the non-linear and log power spectra, via linear combination of indicator spectra (and cross spectra). The fact that one can recover these spectra from indicator spectra reminds us that the passage to indicator functions is invertible and non-destructive of information. It follows that indicator function methods can extract at least as much information as the log transform, which itself yields an essentially sufficient statistic \citep{CarronSzapudi2014}. The importance of characterizing the log spectrum is reflected in the fact that our indicator-function results depend on the ability to predict the log moments $\overline{A}$ and $\sigma_A^2$. Thus the log transform is, via indicator functions, in some ways foundational both to studies of density-dependent clustering and to fitting of CIC data.

Next, because indicator spectra isolate the behavior of specific density regimes, they provide a new way to characterize small-scale redshift-space distortion. Since the strength of the Finger-of-God effect depends on the velocity dispersion within virialized clusters, which in turn depends on the mass of the cluster, it is likely that one could model the effect for each density bin and then linearly combine the indicator spectra for each bin.

In addition, the prescription for indicator-function power spectra allows comparison of observed galaxy indicator spectra with the predicted dark-matter spectra. Such comparison would provide a level-by-level understanding of galaxy bias. The result is additional flexibility in modeling galaxy bias, in that indicator functions naturally handle bias prescriptions that depend non-linearly on the underlying dark matter density (e.g., \citealp{Ising1}). In essence, indicator functions provide a mechanism for convolving the matter spectrum with a tracer bias that is non-linear in delta.

Finally, the joint distribution of Equation~\ref{eq:O2jntprb} applies to a Gaussian $\mathcal{P}(\nu)$. Substitution of the GEV distribution \citep{ReppApdf} for $\mathcal{P}(\nu)$ in this equation could lead to a more accurate expression for the joint distribution $\mathcal{P}(\nu_1,\nu_2)$.

We conclude that indicator functions and their spectra are a promising new tool for analysis of cosmological data. 

\section*{Acknowledgements}
The Millennium Simulation data bases used in this work and the web application providing online access to them were constructed as part of the activities of the German Astrophysical Virtual Observatory (GAVO). IS acknowledges support from National Science Foundation (NSF) award 1616974. We also thank the reviewer for helpful comments and questions.

\section*{Data Availability}
All data used in this work are available from the Millennium Simulation data base at https://wwwmpa.mpa-garching.mpg.de/millennium.

\bibliographystyle{astron}
\bibliography{astro_rsrch}

\appendix
\section{Covariance Matrix Details}
\label{app:cov}

We here provide the details of the covariance matrix calculation for Fig.~\ref{fig:covmatr}, which requires integration of the indicator spectra (and cross spectra) as in Equation~\ref{eq:xibar}. We here explicitly allow for cross spectra by rewriting Equation~\ref{eq:xibar} as follows. If $B_1$ and $B_2$ are two density bins, they define indicator functions $\mathcal{I}_1$ and $\mathcal{I}_2$; if we write the corresponding indicator (cross-)power spectrum as $P_{12}(k)$, we have
\begin{equation}
\overline{\xi}_{12} = \int \frac{d^3k}{(2\pi)^3} P_{12}(k) \prod_{i=1}^3 \left|\, \mathrm{sinc} \left( \frac{k_iL}{2} \right) \right|^2.
\label{eq:appxi}
\end{equation}
If $B_1 = B_2$ then this expression reduces to Equation~\ref{eq:xibar}, and we obtain $\overline{\xi}_\mathcal{I}^{\neq}$ via Equation~\ref{eq:xibarneq}.

We now characterize the (cross) spectra $P_{12}(k)$. We use as our first-order model the biased linear function plus a constant
\begin{equation}
P_{12}(k) = b^2 P_\mathrm{lin}(k) + C,
\label{eq:appord1}
\end{equation}
omitting the power-law term in Equation~\ref{eq:pwrlawmod} -- both because it is not well-characterized and because it becomes important only at larger $k$s, which are excluded by the window function of Equation~\ref{eq:appxi}.

The bias $b^2$ is given by Equations~\ref{eq:biaswrtA} and \ref{eq:1ordpred}; we here combine and rewrite them explicitly in terms of their application to cross-spectra:
\begin{equation}
b^2 = \frac{(\langle A \rangle_{
\!B_1} - \overline{A})(\langle A \rangle_{\!B_2} - \overline{A})}{\sigma_A^4} b_A^2.
\label{eq:bord1}
\end{equation}

The constant $C$ in turn is constrained by the fact that the integral of the indicator spectrum equals the indicator-function variance. If $B_1 = B_2$, we obtain as in Equation~\ref{eq:Clin}
\begin{equation}
C = \delta V \left(\frac{1}{\mathcal{P}(B)} - 1 - b^2 \sigma^2_\mathrm{lin} \right).
\label{eq:Cord1eq}
\end{equation}
On the other hand, if $B_1$ and $B_2$ are disjoint, we have
\begin{align}
\int \frac{d^3k}{(2\pi)^3}P_{12}(k) & = \xi_{12}(0) = \langle \delta_{\mathcal{I}_1}(\mathbf{r}) \delta_{\mathcal{I}_2}(\mathbf{r}) \rangle \\
   & = \left\langle \left( \frac{\mathcal{I}_1}{\mathcal{P}_1} - 1 \right) \left( \frac{\mathcal{I}_2}{\mathcal{P}_2} - 1 \right) \right\rangle \\
   & = -1,\label{eq:intcrspec}
\end{align}
where we write $\mathcal{P}_i = \mathcal{P}(B_i)$ and use the fact that $\langle \mathcal{I}_i \rangle = \mathcal{P}_i$. Then from Equation~\ref{eq:intcrspec} we can derive the analogue of Equation~\ref{eq:Cord1eq} in the case of disjoint bins:
\begin{equation}
C = \delta V \left(-1 - b^2 \sigma^2_\mathrm{lin} \right).
\label{eq:Cord1neq}
\end{equation}

Thus, we use these first-order results (Equations~\ref{eq:appord1}, \ref{eq:bord1}, and either \ref{eq:Cord1eq} or \ref{eq:Cord1neq}) as long as the first-order expansion is valid, namely, when both $| \nu_1 |$ and $|\nu_2|$ are greater than 0.5.

If either $|\nu|$ is less than 0.5, we turn to the second-order expansion, as detailed in Section~\ref{sec:2ordpresc}. Thus our model becomes
\begin{equation}
P_{12}(k) = b_1^2 P_\mathrm{lin}(k) + b_2^2 P_\mathrm{lin}^{(*2)} + C,
\label{eq:appord2}
\end{equation}
with $P_\mathrm{lin}^{(*2)}$ again the self-convolution of the linear power spectrum. The second-order bias is, as in Equation~\ref{eq:2ordpred},
\begin{equation}
b_2^2 = \frac{(1-\langle\nu_1^2\rangle)(1-\langle\nu_2^2\rangle)}{2\sigma_A^4} b_A^4.
\label{eq:bord2}
\end{equation}

To calculate the constant in this case, we must integrate over the self-convolution $P_\mathrm{lin}^{(*2)}$. If we define
\begin{equation}
I_2 = \int \frac{dk}{2\pi^2} P_\mathrm{lin}^{(*2)}(k),
\end{equation}
we obtain the following results: if $B_1 = B_2$,
\begin{equation}
C = \delta V \left(\frac{1}{\mathcal{P}(B)} - 1 - b_1^2 \sigma^2_\mathrm{lin} - b_2^2 I_2 \right);
\label{eq:Cord2eq}
\end{equation}
if $B_1$ and $B_2$ are disjoint,
\begin{equation}
C = \delta V \left(-1 - b_1^2 \sigma^2_\mathrm{lin} - b_2^2 I_2 \right).
\label{eq:Cord2neq}
\end{equation}
Thus, if $|\nu_1|$ or $|\nu_2|$ is less than 0.5, we use these second-order results (Equations~\ref{eq:appord2}, \ref{eq:bord2}, and either \ref{eq:Cord2eq} or \ref{eq:Cord2neq}) in calculating $\overline{\xi}_{12}$.

\label{lastpage}

\end{document}